\documentclass[useAMS,usenatbib,usegraphicx]{mn2e}

\newcommand{\beq}{\begin{equation}}
\newcommand{\eeq}{\end{equation}}
\newcommand{\bdm}{\begin{displaymath}}
\newcommand{\edm}{\end{displaymath}}
\newcommand{\beqa}{\begin{eqnarray}}
\newcommand{\eeqa}{\end{eqnarray}}
\newcommand{\bt}{\begin{tabular}}
\newcommand{\et}{\end{tabular}}

\newcommand{\kv}{{\bf k}}

\newcommand{\qv}{{\bf q}}

\newcommand{\xv}{{\bf x}}

\newcommand{\vv}{{\bf v}}

\def\Mpc{\, h^{-1} \, {\rm Mpc}}

\def\kMpc{\, h \, {\rm Mpc}^{-1}}

\def\fun#1#2{\lower3.6pt\vbox{\baselineskip0pt\lineskip.9pt
        \ialign{$\mathsurround=0pt#1\hfill##\hfil$\crcr#2\crcr\sim\crcr}}}
\def\la{\mathrel{\mathpalette\fun <}}
\def\ga{\mathrel{\mathpalette\fun >}}

\title[Transients from Initial Conditions in Cosmological Simulations]{Transients from Initial Conditions  in Cosmological Simulations}
\author[M. Crocce, S. Pueblas \& R. Scoccimarro]{Mart\'{\i}n Crocce, Sebasti\'an Pueblas and Rom\'an Scoccimarro\\
Center for Cosmology and Particle Physics, Department of Physics, New York University, New York, NY 10003}
\begin{document}

%\date{}

\pagerange{\pageref{firstpage}--\pageref{lastpage}} \pubyear{2006}

\maketitle

\label{firstpage}

\begin{abstract}

We study the impact of setting initial conditions in numerical simulations using the standard procedure based on the Zel'dovich approximation (ZA). As it is well known from perturbation theory, ZA initial conditions have incorrect second and higher-order growth and therefore excite long-lived transients in the evolution of the statistical properties of density and velocity fields. We also study the improvement brought by using more accurate initial conditions based on second-order Lagrangian perturbation theory (2LPT). We show that 2LPT initial conditions reduce transients significantly and thus are much more appropriate for numerical simulations devoted to precision cosmology. Using controlled numerical experiments with ZA and 2LPT initial conditions we show that simulations started at redshift $z_i=49$ using the ZA underestimate the power spectrum  in the nonlinear regime by about $2,4,8\%$ at $z=0,1,3$ respectively, whereas the mass function of dark matter halos is underestimated by $5\%$ at $m=10^{15}M_\odot/h$ ($z=0$) and $10\%$ at $m=2\times 10^{14}M_\odot/h$ ($z=1$). The clustering of halos is also affected to the few percent level at $z=0$. These systematics effects are typically larger than statistical uncertainties in recent mass function and power spectrum fitting formulae extracted from numerical simulations.
At large scales, the measured transients in higher-order correlations can be understood from first principle calculations based on perturbation theory.

\end{abstract}


\section{Introduction}

The growth of structure in the universe is a fundamental aspect of cosmology, and can be used to constrain the nature of primordial fluctuations and the energy contents of the universe (e.g. dark matter and dark energy). Numerical simulations are a very important tool in making theoretical predictions of cosmological observables  at small scales when fluctuations are nonlinear. Fluctuations are followed from their primordial seeds  through the radiation era to the matter dominated era and current accelerating phase by using linear Boltzmann equation solvers \citep{PeYu70,BE84,MaBe95,CMBFAST,CAMB}, which are  then coupled to N-body simulations to follow their nonlinear evolution. This coupling takes the form of final conditions predicted from linear Boltzmann codes input as initial conditions into N-body simulations at some redshift $z_i$. 

There are competing requirements on the best choice of $z_i$. One would like to start a simulation as early as possible, where perturbations are closest to their linear value but this can be challenging from numerical reasons since one is trying to impose a small perturbation spectrum that can be overwhelmed by sources of noise such as numerical roundoff error or shot noise of the particles. Also, when running simulations that include a baryonic component, the initial conditions simplify considerably the lower $z_i$ is, in which case the baryons had more time to fall into the potential wells of the dark matter. 

Despite the rapid progress in the development of different numerical N-body techniques for solving the nonlinear growth of structure in the universe (see e.g. \citet{Bert98} for a review), the way initial conditions  are input into these codes has essentially been the same since the first developments in cosmological simulations over twenty years ago \citep{KlSh83,EDFW85}, by using the \citet{ZA70} approximation (hereafter ZA).

As observations of the late universe quickly approach percent-level accuracy, it is vital to examine to what extent theoretical predictions can accurately go from primordial fluctuations to different stages of nonlinear evolution in recent epochs, e.g. redshifts $z \la 3$. Studies of linear Boltzmann solvers \citep{CMBcodes} have shown that independent codes agree to within 0.1\%, whereas for N-body codes the agreement, while much more difficult to quantify, is of the order of few to ten percent depending on the statistics and scales considered \citep{SMSS98,SBcluster,HRWH05}.

The link between Boltzmann and N-body codes, or how to set initial conditions for nonlinear evolution, has been surprisingly little studied. The accuracy of the ZA has been essentially taken for granted and instead studies of initial conditions have concentrated on other (important) issues, related to how to best implement the ZA in different situations \citep{Bert95,Salmon96,Pen97,Bert01,YSH03,Sirko05}. 

The rationale for this state of affairs seems to be that starting a simulation at  ``high enough" redshift with linear perturbation theory (which the ZA is, in Lagrangian space) should be safe. A rule of thumb for the choice of starting redshift often used is that the {\em rms} density fluctuations at the interparticle distance scale be of order $0.2$. As a result, low resolution simulations are typically started much later than high resolution ones.

However, one of the lessons of perturbation theory (hereafter PT, see \cite{PTreview} for a review) is that the ZA has a very limited validity, reproducing only the correct linear growth and underestimating the skewness and higher-order moments \citep{GW87,JBC93,B94,CM94,J95,BGE95,CBH96}, with deviations being particularly severe for velocity field perturbations because of the ZA failure to conserve momentum.

As a result of this, starting N-body simulations from ZA initial conditions leads to incorrect second and higher-order growing modes; this excites nonlinear decaying modes (called {\em transients}) that describe how long it takes for the correct dynamics to recover the actual statistical properties of density and velocity fields  \citep{Sco98}. This transient evolution is well understood at large scales using PT and describes  the anomalous behavior seen in measurements of higher-order moments in N-body simulations started from ZA initial conditions (as shown in \citet{Sco98} and demonstrated in much more detail below).  

The importance of transients is not just of interest for studies involving higher-order moments, as we show in this paper. Underestimation of  non-Gaussianity means that the tails of the probability distribution of densities and velocities will be artificially suppressed, which implies that rare events will be even more rare in a simulation that has transients. Since typically large volumes (and thus typically low resolution) are used to study rare events, the dark matter halo mass function at the high-mass end will be particularly affected by transients given that low-resolution simulations are often started later than high-resolution ones.   In addition, the transient behavior of large-scale higher-order moments implies that nonlinear couplings take a significant time to develop their full strength. These are the same couplings that act at small scales and give rise to the nonlinear evolution of the power spectrum, therefore transients should also affect the power spectrum at nonlinear scales.

In this paper we revisit the issue of transients from ZA initial conditions and study numerically the effects on higher-order moments, the mass function of dark matter halos and the power spectrum. As proposed in \citet{Sco98}, transients can be reduced significantly by using second-order Lagrangian PT (hereafter 2LPT). This leads to correct second-order growth factors and greatly improves the behavior of higher-order moments. Furthermore, transients in this case decay as inverse square power of the scale factor, much faster than the inverse of the scale factor in the ZA case. We study the improvement brought by using 2LPT instead of ZA in the behavior of transients and quantify their magnitude in different statistics commonly used.

The paper is organized as follows. In section~\ref{basics} we review how transients arise and present predictions that are contrasted with simulations later. Section~\ref{simu} discusses the numerical simulations we run to quantify the transients in the nonlinear regime. In section~\ref{Res} we present the main results of the paper on the magnitude of transients in higher-order cumulants, power spectrum, bispectrum, mass function and bias of dark matter halos. We summarize the results and conclude in section~\ref{Conc}. In appendix~\ref{sec2LPT} we discuss second-order Lagrangian perturbation theory and in appendix~\ref{secGlass} we derive analytically the evolution of perturbations toward a glass configuration and discuss glass initial conditions from the point of view of transients.

\section{Transients}
\label{basics}

In this section we discuss why transients exist and how they affect the statistics of density and velocity fields at large scales, following the treatment by \citet{Sco98}, to which we refer the reader for more details. The analytic results summarized here also give a framework to understand the measurements in simulations we discuss below at small scales. Throughout this paper we assume Gaussian primordial  fluctuations. For the most part we assume a flat universe with $\Omega_m=1$ and $\Omega_{\Lambda}=0$, we mention the more general case at the end of section~\ref{EoM}.

\subsection{Equations of Motion}
\label{EoM}

The equations of motion for the evolution of dark matter can be written in a compact way by introducing the two-component ``vector''

\beq
\Psi_a(\kv,\eta) \equiv \Big( \delta(\kv,\eta),\ -\theta(\kv,\eta)/{\cal H} \Big),
\label{2vector}
\eeq
where the index $a=1,2$ selects the density or velocity components, with $\delta(\kv)$ being the Fourier transform of the density contrast  $\delta (\xv,\tau)=\rho(\xv)/\bar \rho - 1$ and similarly for the peculiar velocity divergence $\theta\equiv\nabla\cdot{\bf v}$. ${\cal H}\equiv {d\ln a /{d\tau}}$ is the conformal expansion rate 
with  $a(\tau)$ the cosmological scale factor and $\tau$ the conformal time. The time variable $\eta$ is defined from the scale factor by 

\beq
\eta\equiv\ln a(\tau),
\label{timevariable}
\eeq
corresponding to the number of e-folds of expansion.  The equations of motion in Fourier space can then be written as (we henceforth use the convention that repeated Fourier arguments are integrated over)
\beq
\partial_{\eta} \Psi_a(\kv) + \Omega_{ab} \Psi_b(\kv) =
 \gamma_{abc}^{(\rm s)}(\kv,\kv_1,\kv_2)\  \Psi_b(\kv_1)\ \Psi_c(\kv_2), 
\label{eom}
\eeq
where 
\beq
\label{Omab}
\Omega_{ab} \equiv \Bigg[ 
\begin{array}{cc}
0 & -1 \\ -3/2 & 1/2 
\end{array}        \Bigg],
\eeq
describes the linear structure of the equations of motion. In particular, the first row describes the linearized continuity equation $\partial_\eta \Psi_1-\Psi_2=0$ and the second row describes linearized momentum conservation, with the two entries being the Poisson equation ($\nabla^2 \Phi =(3/2) {\cal H}^2 \delta$) and the Hubble drag term, respectively. The symmetrized vertex matrix $\gamma_{abc}^{(\rm s)}$ describes the non linear interactions between different Fourier modes and is given by

\beqa
\gamma_{121}^{({\rm s})}(\kv,\kv_1,\kv_2)&=&\delta_{\rm D}(\kv-\kv_1-\kv_2) \  {(\kv_1+\kv_2) \cdot \kv_1 \over{2 k_1^2}},
\label{ga121} \\
\gamma_{222}^{({\rm s})}(\kv,\kv_1,\kv_2)&=&\delta_{\rm D}(\kv-\kv_1-\kv_2) \ {|\kv_1+\kv_2|^2 (\kv_1 \cdot \kv_2 )\over{2 k_1^2 k_2^2}},
\nonumber \\ & & 
\label{ga222}
\label{vertexdefinition}
\eeqa
$\gamma_{abc}^{(\rm s)}(\kv,\kv_i,\kv_j)=\gamma_{acb}^{(\rm s)}(\kv,\kv_j,\kv_i)$, and $\gamma$ is zero otherwise, $\delta_{\rm D}$ denotes the Dirac delta distribution. The formal integral solution to Eq. (\ref{eom}) is given by (see \citet{Sco00} for a detailed derivation)

\beqa
\Psi_a(\kv,\eta) &=& g_{ab}(\eta) \ \phi_b(\kv) + \int_0^\eta  d\eta' \ g_{ab}(\eta-\eta')
\nonumber \\ & & \times  \gamma_{bcd}^{(\rm s)}(\kv,\kv_1,\kv_2)
 \Psi_c(\kv_1,\eta') \Psi_d(\kv_2,\eta'),
\label{eomi}
\eeqa
where  $\phi_a(\kv)\equiv\Psi_a(\kv,\eta=0)$ denotes the initial conditions, set when the linear growth factor $a(\tau)=1$ and $\eta=0$. The {\em linear propagator} $g_{ab}(\eta)$ is simply related to $\Omega_{ab}$, and given by an inverse Laplace transform ($\eta\geq 0$)

\beq
g_{ab}(\eta) = \oint \frac{ds}{2\pi i}\ {\rm e}^{s\eta}\ (\Omega_{ab}+s\,\delta_{ab})^{-1}
\label{propdef}
\eeq
which given Eq.~(\ref{Omab}) yields
\beq
g_{ab}(\eta) =\frac{{\rm e}^{\eta}}{5}
\Bigg[ \begin{array}{rr} 3 & 2 \\ 3 & 2 \end{array} \Bigg] -
\frac{{\rm e}^{-3\eta/2}}{5}
\Bigg[ \begin{array}{rr} -2 & 2 \\ 3 & -3 \end{array} \Bigg].
\label{prop}
\eeq 
This describes the standard linear evolution of the density and velocity fields from their initial conditions specified by   $\phi_a(\kv)$. 

In the ZA, the Poisson equation is replaced by $\nabla^2 \Phi =- (3/2) {\cal H}\ \theta$ \citep{MuSt94,HuBe96}, and thus Eq.~(\ref{Omab}) is replaced by,
\beq
\label{OmabZA}
\Omega_{ab}^{ZA} \equiv \Bigg[ 
\begin{array}{cc}
0 & -1 \\ 0 & -1 
\end{array}        \Bigg],
\eeq
which from Eq.~(\ref{propdef}) gives the linear propagator
\beq
g_{ab}^{ZA}(\eta) = {\rm e}^{\eta}
\Bigg[ \begin{array}{rr} 0 & 1 \\ 0 & 1 \end{array} \Bigg] +\Bigg[ \begin{array}{rr} 1 & -1 \\ 0 & 0 \end{array} \Bigg].
\label{propZA}
\eeq 
The change in the linear propagator (a reflection of gravity being driven by the velocity divergence instead of density fluctuations) has important consequences, as its structure determines the generation of non-Gaussianity.

From Eq.~(\ref{eomi}) we can see why transients arise. For simplicity, let's assume first that the equations of motion are linear, i.e. that $\gamma_{abc}=0$ and the second term in the right hand side of Eq.~(\ref{eomi}) is zero. Then, selecting growing mode initial conditions is very simple. All we have to do is to select $\phi_a(\kv)=\delta_0(\kv)\ u_a$, where $\delta_0(\kv)$ is a Gaussian random field whose power spectrum follows from a Boltzmann solver and $u_a=(1,1)$ selects the growing mode upon contraction with $g_{ab}$ in Eq.~(\ref{prop}). Selecting the growing mode is easy since we can solve the linear evolution analytically (given by the first term in Eq.~(\ref{eomi})), otherwise one would generally excite the decaying mode.

In the nonlinear case, we do not know an exact solution (that's the purpose of running a simulation), therefore we cannot select initial conditions in the exact (nonlinear) growing mode. The best we can do is to try to use PT to select the fastest growing mode at each order in PT (2LPT achieves this to second-order, and does a good job for higher orders as well).  Since the initial conditions will not be in the exact growing mode, decaying modes will be excited by nonlinear interactions in the second term of Eq.~(\ref{eomi}): these are what we call {\em transients}. This is analogous to what would happen in the linear case if  $\phi_a(\kv)$ was not the eigenvector $(1,1)$, i.e. the decaying mode in the second term of Eq.(\ref{prop}) would be excited.

In the nonlinear case the problem of transients is worse than one might think based on linear theory because transients do not go away as fast as in linear theory, where the decaying mode is suppresed by $a^{-5/2}$ compared to the growing mode. Indeed, if the ZA is used to set initial conditions transients are only suppressed by $a^{-1}$, so they take a long time to decay. Using better initial conditions, such as 2LPT, leads to $a^{-2}$ suppression, a substantial improvement.

To describe transients in a quantitative way, we expand the initial and final conditions in terms of Gaussian linearly evolved density perturbations  $\delta_0(\kv)$,

\beq
\phi_a(\kv)=\sum_{n=1}^{\infty}\, \phi^{(n)}_a(\kv), \ \ \ \ \ \Psi_a(\kv,\eta)=\sum_{n=1}^{\infty}\,  \Psi^{(n)}_a(\kv,\eta),
\label{seriesexp}
\eeq
with 
\beq
\phi_a^{(n)}(\kv)=[\delta_{\rm D}]_n\ {\cal I}_{a}^{(n)}(\kv_1,\ldots,\kv_n) \ \delta_0(\kv_1) \ldots \delta_0(\kv_n), 
\label{seriesolIC}
\eeq
\beq
\Psi_a^{(n)}(\kv,\eta)=[\delta_{\rm D}]_n\ {\cal F}_{a}^{(n)}(\kv_1,\ldots,\kv_n;\eta) \ \delta_0(\kv_1) \ldots \delta_0(\kv_n), 
\label{seriesol}
\eeq
where $[\delta_{\rm D}]_n \equiv  \delta_{\rm D}(\kv-\kv_{1\ldots n})$ and $\kv_{1\ldots n} \equiv \kv_1+ \ldots + \kv_n$. The kernels ${\cal I}_a^{(n)}$ describe how the initial conditions are set (at $\eta=0$), e.g. if linear theory is used ${\cal I}_a^{(1)}=u_a=(1,1)$ and ${\cal I}_a^{(n)}=0$ for $n>1$. The kernels ${\cal F}_a^{(n)}(\eta)$ describe how perturbations evolve after initial conditions are set, they obey the recursion relation \citep{Sco98},

\beqa
 {\cal F}_a^{(n)}(\eta)  &=&  \sum_{m=1}^{n-1} \int_0^\eta ds \,g_{ab}(\eta-s)  \gamma_{bcd}^{(\rm s)}
  \, {\cal F}_c^{(m)}(s) \, {\cal F}_d^{(n-m)}(s) \nonumber \\ & & +\ g_{ab}(\eta)\ {\cal I}_b^{(n)},
\label{recursionkernels}
\eeqa
obtained by replacing Eq.~(\ref{seriesexp}) and  Eqs.~(\ref{seriesolIC}-\ref{seriesol}) into  Eq.~(\ref{eomi}). For brevity we have suppressed the Fourier arguments. Note that  ${\cal F}_a^{(n)}(\eta=0)={\cal I}_a^{(n)}$ as it should be, afterwards the fastest growing mode behaves as ${\cal F}_a^{(n)}(\eta) \sim {\rm e}^{n \eta}$ at each order $n$. {\em Transients determine how fast one goes from initial conditions given by ${\cal I}_a^{(n)}$ to the regime where final conditions are in the fastest growing mode ${\cal F}_a^{(n)}(\eta) \sim {\rm e}^{n \eta}$}. 

If the initial conditions are not set using linear theory, the kernels ${\cal I}_a^{(n)}$ obey recursion relations that follow from the dynamics used to set them. For ZA initial conditions ${\cal I}_a^{(n)}$ obeys,

\beqa
 {\cal I}_a^{(n)} &=&  \sum_{m=1}^{n-1} \int_{-\infty}^\eta ds \, {\rm e}^{n(s-\eta)}g_{ab}^{ZA}(\eta-s)  \gamma_{bcd}^{(\rm s)}
  \, {\cal I}_c^{(m)} \, {\cal I}_d^{(n-m)}, \nonumber \\ & & 
\label{recursionkernelsZA}
\eeqa
with the condition ${\cal I}_a^{(1)}=(1,1)$. This can be understood from Eq.~(\ref{recursionkernels}) by using linear theory initial conditions and keeping the fastest growing mode, formally done by evolving by an infinite amount of time (from $s=-\infty$). Note that these kernels are independent of $\eta$, that's guaranteed due to the lower limit in the time integral. In the long time limit, the ${\cal F}_a^{(n)}(\eta)$ kernels  approach ${\cal F}_a^{(n)}(\eta) = {\rm e}^{n \eta}\, {\cal H}_a^{(n)}$, where ${\cal H}_a^{(n)}$ obeys the recursion relation in Eq.~(\ref{recursionkernelsZA}) with the linear propagator $g_{ab}^{ZA}$ replaced by $g_{ab}$, Eq.~(\ref{prop}). It is easy to check that if ${\cal I}_a^{(n)}$ is replaced by such ${\cal H}_a^{(n)}$ in Eq.~(\ref{recursionkernels}), then ${\cal F}_a^{(n)}(\eta) = {\rm e}^{n \eta}\, {\cal H}_a^{(n)}$ is a solution at all times, i.e. there are no transients, as expected since one is setting initial conditions exactly, i.e. in the fastest growing mode.

Although most simulations are started by initial conditions decribed by the kernels in Eq.~(\ref{recursionkernelsZA}) when using the ZA, some authors \citep{EDFW85,Petal03} define the velocities in a different way, by applying the ZA prescription to the {\em perturbed} density field after particles have been moved instead of the linear density field. In this case, it can be shown \citep{Sco98} that the velocity divergence kernels are not those in the ZA, but rather they are equal to the {\em density} kernels in the ZA. That leads to a reduction in the time scale of transients by a factor of about two \citep{Sco98}. We will not consider this possibility here, as the standard ZA starts are most often used in the literature, and the improvement brought by using this prescription for the velocities does not fully address the limitations of the ZA (e.g. the second-order growth is incorrect, and the transients still evolve as the inverse of the scale factor).

As an example, the second-order density kernel reads from Eq.~(\ref{recursionkernels}), with $x \equiv \hat{k}_1 \cdot \hat{k}_2$, 

\beqa
{\cal F}_1^{(2)}(\eta)& = & {\rm e}^{2\eta} \Big[ \frac{5}{7} +\frac{x}{2}\Big( \frac{k_1}{k_2} + \frac{k_2}{k_1} \Big) + \frac{2}{7} x^2 \Big]
+{\rm e}^{\eta} \frac{3}{10} (x^2-1) \nonumber \\ & & +\ {\rm e}^{-3\eta/2}\, \frac{3 }{35} (x^2-1),
\label{F2trans}
\eeqa
which interpolates between the ZA value at $\eta=0$ and the exact value for large $\eta$ (that in between square brackets). Note that, as mentioned above, the dominant transient term is suppressed only by a single inverse power of the scale factor, so it takes a long time to disappear. If 2LPT is used to set initial conditions, the second-order kernels are reproduced exactly and there are no transients to this order. For higher-order kernels, 2LPT shows transients suppressed by inverse square powers of the scale factor \citep{Sco98}, and as we see below they are greatly reduced compared to the ZA case.

In making predictions that evolve all the way to $z=0$ one must take into account the dependence on cosmological parameters neglected so far. A simple and rather accurate way of doing this is to redefine in Eq.~(\ref{2vector}), $ -\theta(\kv,\eta)/f{\cal H} $ as the second component, where $f \equiv d\ln D_+/d\ln a$ and $D_+$ is the growth factor with the exact cosmological dependence. When this is done, the only change in the equations of motion is $3/2 \rightarrow 3\Omega_m/2f^2$ in Eq.~(\ref{Omab}), and given that $f \simeq \Omega_m^{5/9}$ \citep{BCHJ95} for flat models with a cosmological constant this can be well approximated by the $\Omega_m=1$ value of $3/2$ since during most of the time evolution $\Omega_m/f^2$ is very close to unity. This leads to the same kernels as in the $\Omega_m=1$ case but with the scale factor replaced everywhere by the growth factor $D_+$. 

\subsection{Statistics}
\label{Stat}

We now discuss quantitatively the impact of transients on statistics of the density field at large scales. Since linear growing modes are reproduced exactly, the power spectrum at large scales is not affected by transients. The effect of transients a large scales is only felt by higher-order spectra, such as the bispectrum, or cumulants, e.g. the skewness and kurtosis. However, as these quantities characterize non-linear couplings, they are useful for understanding the results at small scales where nonlinearities play a role in the power spectrum, which will then be affected by transients, as we shall see using simulations in section~\ref{secPower}.

The power spectrum $P(k)$, bispectrum $B_{123}\equiv B(k_1,k_2,k_3)$ and trispectrum $T_{1234}$ are given by,
\beqa
\langle \delta(\kv) \delta(\kv') \rangle &=& \delta_{\rm D}(\kv+\kv')\ P(k), \\ 
\langle \delta(\kv_1) \delta(\kv_2) \delta(\kv_3) \rangle &=& \delta_{\rm D}(\kv_{123})\ B_{123},  \\
\langle \delta(\kv_1) \delta(\kv_2) \delta(\kv_3) \delta(\kv_4)\rangle_c &=& \delta_{\rm D}(\kv_{1234})\ T_{1234},  
\eeqa
with $\kv_{123}\equiv \kv_1+\kv_2+\kv_3$, etc., and the subscript ``c" denotes a connected contribution, whereas the skewness $S_3$ and kurtosis $S_4$ are given by,
\beq
S_3=\frac{\langle \delta^3 \rangle}{\langle \delta^2 \rangle^2}, \ \ \ \ \ \ \ \ \ \ 
S_4=\frac{\langle \delta^4 \rangle_c}{\langle \delta^2 \rangle^3}.
\eeq
In second-order perturbation theory the bispectrum reads

\beq
B_{123} = 2 {\cal F}_1^{(2)}(\kv_1,\kv_2;\eta)\ P_0(k_1)\, P_0(k_2) +~{\rm cyclic},
\label{Bisp2E}
\eeq
where $P_0$ denotes the power spectrum of the linear fluctuations at the initial conditions (i.e. $P(k)=a^2 P_0(k)$ in linear theory), and cyclic permutations over $\{k_1,k_2,k_3\}$ are understood. The first three cumulants are given by,

\beq
\label{mom2}
\langle \delta^2 \rangle =  \int   P(k)\, W^2(kR)\, d^3k,
\eeq
\beq
\label{mom3}
\langle \delta^3 \rangle =  \int B_{123} \, W_1W_2 W_3 \, \delta_{\rm D}(\kv_{123})\, d^3 k_1 d^3 k_2 d^3 k_3,
\eeq
\beq
\label{mom4}
\langle \delta^4 \rangle_c =  \int T_{1234} \, W_1W_2 W_3 W_4\, \delta_{\rm D}(\kv_{123})\, d^3 k_1 \ldots  d^3 k_4
\eeq
where $W(kR)$ represents a top hat filter in Fourier space at smoothing scale $R$, and $W_i \equiv W(k_iR)$. 
From these expressions it follows the skewness and kurtosis including transients \citep{Sco98}

\beqa
S_3 &=& \Big\{ \frac{34}{7} -\gamma_1 \Big\} + \frac{1}{a} [ 4 -\gamma_1 ]  + \frac{1}{a} \Big( \gamma_1 -\frac{26}{5} \Big) + \frac{12}{35 a^{7/2}}, \nonumber \\ &&
\label{S3trans}
\eeqa
\beqa
S_4 &=& \frac{60712}{1323} -\frac{62\gamma_1}{3}+ \frac{7\gamma_1^3}{3}-\frac{2\gamma_2}{3}-\frac{816}{35a}+\frac{28\gamma_1}{5a}
+ \frac{184}{75 a^2}
\nonumber \\  &&  + \frac{1312}{245 a^{7/2}} - \frac{8 \gamma_1}{5 a^{7/2}} -\frac{1504}{4725 a^{9/2}} +\frac{192}{1225 a^7},
\label{S4trans}
\eeqa

where we have written the time dependence in terms of the scale factor $a$ and

\beq
\gamma_p \equiv - \frac{d^p \ln \sigma^2(R)}{d \ln^p R},
\eeq
with $\sigma^2(R)$ the variance of density perturbations, Eq.~(\ref{mom2}).

These results have also been derived more recently using a different approach based on the steepest descent method for the density PDF  \citep{Val02}. In Eq.~(\ref{S3trans}) we have explicitly written down the different contributions: the ZA contribution from initial conditions in square brackets, the exact value (reached only asymptotically as $a\rightarrow \infty$) in curly brackets, and the remaining two transient terms (which cancel the exact value when $a=1$). If initial conditions were set using Eulerian linear theory, the term in square brackets would not be present.

\begin{figure}
\includegraphics[width=84mm]{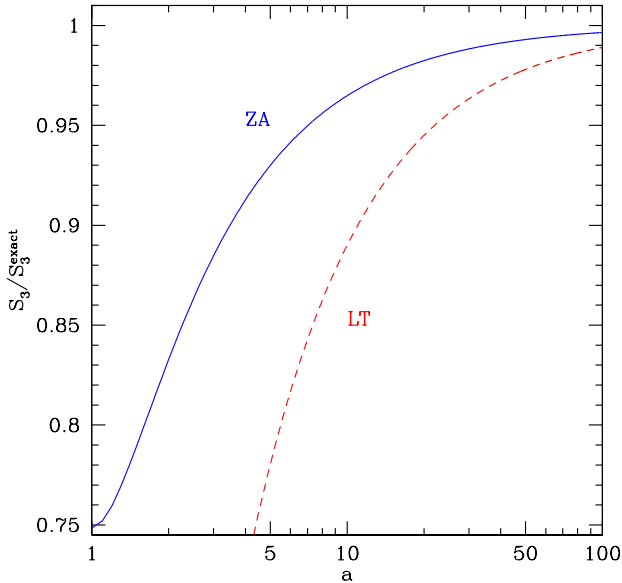}
\caption{The skewness $S_3$ at $R=10 \Mpc$ for Zel'dovich approximation (solid) and Eulerian linear theory (dashed) initial conditions as a function of the expansion factor $a$ away from initial conditions (at $a=1$). Values are given in terms of the exact value of $S_3$. }
\label{S3ZALT}
\end{figure}

As discussed above, to include the dependence of transients on cosmological parameters we will make the replacement $a \rightarrow D_+$ in all comparisons of Eqs.~(\ref{S3trans}-\ref{S4trans}) with numerical simulatons.

Figure~\ref{S3ZALT} evaluates Eq.~(\ref{S3trans}) in two cases, one where ZA is used to set the initial conditions (solid), and the other when Eulerian linear theory is used (and thus $S_3=0$ at $a=1$). We see that if linear theory were used, to reach the correct value of $S_3$ within some tolerance, it would take about four times more expansion away from the initial conditions compared to the ZA case. The difference in the two cases is purely due to the initial value of $S_3$, i.e. the rate of convergence to the right answer is the same (as transients  go away as $a^{-1}$ in both cases). In contrast, using 2LPT initial conditions improves the initial values of the $S_p$ parameters dramatically over the ZA case (being exact for $S_3$), and also improves substantially the rate of decay of transients to $a^{-2}$.

Linear theory initial conditions are relevant for codes that do no use particles, e.g. in the case where baryons are described by fields in a grid instead of particles. In such cases one must keep in mind that the magnitude of the transients we demonstrate with simulations below are very likely to be much worse, as reflected by the difference between the curves in Fig.~\ref{S3ZALT}. 

The dependence of transients on the shape of the power spectrum, in Eqs.~(\ref{S3trans}-\ref{S4trans}) through $\gamma_p$, shows that transients from ZA initial conditions are somewhat reduced as the spectrum becomes steep, since the gravitational interactions are dominated by large scales and this leads to fewer deflections, i.e. more coherent displacement fields better approximated by straight-line trajectories given by the ZA.  The dependence on spectrum is discussed in detail in \citet{Sco98}. Here we concentrate on a fixed shape given by a concordance cosmological model as discussed in the next section.

The lesson from Fig.~\ref{S3ZALT} is that nonlinear couplings (in this case, second order) take a significant time to develop their full strength if initial conditions are not set accurately. As expected from perturbation theory, these weaker nonlinear couplings give a transient effect in the nonlinear power spectrum. The effect of transients is also felt rather strongly on rare events such as the abundance of high-mass clusters, which probe the tail of the density probability distribution characterized by the $S_p$ parameters. We demonstrate these effects below using numerical simulations.

\begin{table*}
 \centering
 \begin{minipage}{150mm}
  \caption{Numerical simulations parameters. All our simulations have $N_{\rm par}=512^3$ particles with $\Omega_m=0.27$, $\Omega_\Lambda=0.73$, $\Omega_b=0.046$, $h=0.72$ and $\sigma_8(z=0)=0.9$, were run using {\sf Gadget2} \citep{Gadget2} with time integration accuracy parameters {\tt ErrTolIntAccuracy}=0.025, {\tt MaxRMSDisplacementFac}=0.2, {\tt MaxSizeTimestep}=0.025, and gravitational force accuracy parameters {\tt ErrTolTheta}=0.5, {\tt TypeOfOpeningCriterion}=1, {\tt ErrTolForceAcc}=0.005 except for the HQ runs, where more stringent parameters ({\tt ErrTolIntAccuracy}=0.01, {\tt MaxRMSDisplacementFac}=0.1, {\tt MaxSizeTimestep}=0.01, {\tt ErrTolTheta}=0.2, {\tt ErrTolForceAcc}=0.002) were chosen. Length units are in $\Mpc$ and mass units are in $10^{10}M_\odot/h$.}
\begin{tabular}{@{}lcccrccr@{}}  \hline
Name  &  $L_{\rm box}$ & $m_{\rm par}$& Initial Conditions & $z_i$ & $N_{\rm realizations}$ & softening & $N_{\rm timesteps}$  \\
 \hline
LRZA49 & 1024 & 59.94 & ZA      & 49     & 8 & 0.2   & 800    \\
LRZA24 & 1024 & 59.94 & ZA      & 24     & 8 & 0.2   & 750    \\
LR2LPT49\footnote{This is the reference run for  $L_{\rm box}=1024$.} & 1024 & 59.94 & 2LPT & 49     & 8 & 0.2   & 800    \\
LR2LPT11.5 & 1024 & 59.94 & 2LPT & 11.5  & 8 & 0.2   & 650    \\
LRsZA49 & 1024 & 59.94 & ZA      & 49     & 2 & 0.07 & 1800    \\
LRs2LPT49 & 1024 & 59.94 & 2LPT & 49     & 2 & 0.07  & 1800    \\
HRZA49 & 512   &   7.49 & ZA      & 49     & 4 & 0.04 & 2200  \\
HRZA24 & 512   &   7.49 & ZA      & 24     & 4 & 0.04 & 2200  \\
HR2LPT49\footnote{This is the reference run for  $L_{\rm box}=512$.} & 512   &   7.49 & 2LPT      & 49     & 20 & 0.04 & 2200  \\
HR2LPT11.5 & 512   &   7.49 & 2LPT      & 11.5     & 4 & 0.04 & 2200  \\
HQZA49 & 512   &   7.49 & ZA      & 49     & 1 & 0.02 & 6500  \\
HQ2LPT49 & 512   &   7.49 & 2LPT      & 49     & 1 & 0.02 & 6500  \\
\hline
\label{SimTable}
\end{tabular}
\end{minipage}
\end{table*}

\section{Simulations}
\label{simu}

In order to study the effects of transients at small scales we ran a set of simulations with different initial conditions (ZA and 2LPT), starting redshifts $z_i$, box sizes $L_{\rm box}$, softening lengths and number of timesteps as described in Table~\ref{SimTable}. All our simulations have $N_{\rm par}=512^3$ particles with cosmological parameters  $\Omega_m=0.27$, $\Omega_\Lambda=0.73$, $\Omega_b=0.046$, $h=0.72$ and $\sigma_8(z=0)=0.9$, and were run using the {\sf Gadget2} code \citep{Gadget2}.

All the internal parameters to the  {\sf Gadget2} code used to run the different numerical experiments are also given in Table~\ref{SimTable}. In summary, we have two sets of initial conditions (ZA and 2LPT), each of them started at two different redshifts $z_i$ with the purpose of checking the magnitude of transients. The difference between ZA and 2LPT initial conditions is explained in Appendix~\ref{sec2LPT} in terms of Lagrangian displacements. In our implementation, particles are started from a grid and all differentiations are done in Fourier space, i.e. through $\nabla \rightarrow i \kv$. Another possibility would be to start particles from a ``glass", in appendix~\ref{secGlass} we briefly discuss how to avoid inducing extra initial skewness (and thus transients) due to glass initial conditions.

A simulation without transients should have statistical properties (measured at redshift $z<z_i$) independent of $z_i$. Also, if generating initial conditions with the ZA was accurate enough, certainly 2LPT would be as well, since PT converges very well at high redshift. If this were the case, no appreciable difference would be seen at low redshift between ZA and 2LPT initial conditions.

However, for ZA initial conditions, we observe that statistics at low redshift {\em are different} than those from 2LPT initial conditions started at the same redshift (in this case $z_i=49$), {\em and} we also see a clear dependence on $z_i$ (comparing $z_i=49$ with $z_i=24$ simulations). Furthermore, the magnitude of transients observed at large scales agrees with the perturbative predictions presented in the previous section. In addition, the expected suppression of nonlinear couplings at large scales acts at small scales in a manner consistent with what one expects physically.

How do we know the 2LPT initial conditions simulations are correct? At large scales we checked the numerical simulations against the PT predictions, and we see almost no dependence of the low redshift results on the starting redshift $z_i$. To make this latter point more  obvious, we decided to present results for a {\em very late start}, $z_i=11.5$ (compared to our standard choice $z_i=49$). {\em Note we are not advocating starting simulations at $z_i=11.5$, this is only to prove that 2LPT initial conditions have very little transient effects}. 2LPT initial conditions {\em do have} transient effects, although much smaller than ZA. Starting at redshift $z_i=11.5$ makes them large enough to be measurable at $z=3$, which allows us to estimate the magnitude of transients in our reference runs (see section~\ref{Conc} for a discussion of this). 

Most of the results presented here are obtained by using the ``high-resolution" runs (labeled HR in Table~\ref{SimTable}), corresponding to $L_{\rm box}=512 \Mpc$. The ``low-resolution"  (labeled LR) simulations with  $L_{\rm box}=1024 \Mpc$ are used to study the high-mass tail of the dark matter halo mass function in Section~\ref{secMF}. In addition, we have studied if the magnitude of transients depends on the softening length by reducing the softening length by a factor of three in the LR runs (denoted by LRs), and we have also run HR simulations with twice smaller softening length and stricter time-integration and force accuracy parameters, labeled ``high-quality" (HQ) runs  in Table~\ref{SimTable}. From these sanity checks we conclude that the magnitude of transients observed from ZA initial conditions we present below is insensitive to the choices we have made in the less time-consuming  runs.

The runs with 2LPT initial conditions with $z_i=49$ are our most accurate simulations from the point of view of transients.  We denote such realizations as {\em reference runs}. In order to quantify the magnitude of transients, we measure statistics at low redshift and compare them to the same measurements in these reference runs.

\section{Results}
\label{Res}

%It was comparing RPT to simulations that led us to consider transients, this exemplifies the usefulness of having analytic solutions to the growth of structure.

\subsection{Higher-Order Cumulants}
\label{secMom}

We first start by comparing simulations for measurements of higher-order cumulants at large scales, skewness $S_3$ and kurtosis $S_4$, which are sensitive to transients and can be understood analytically using perturbation theory, as presented in Section~\ref{Stat}.

\begin{figure}
\includegraphics[width=84mm]{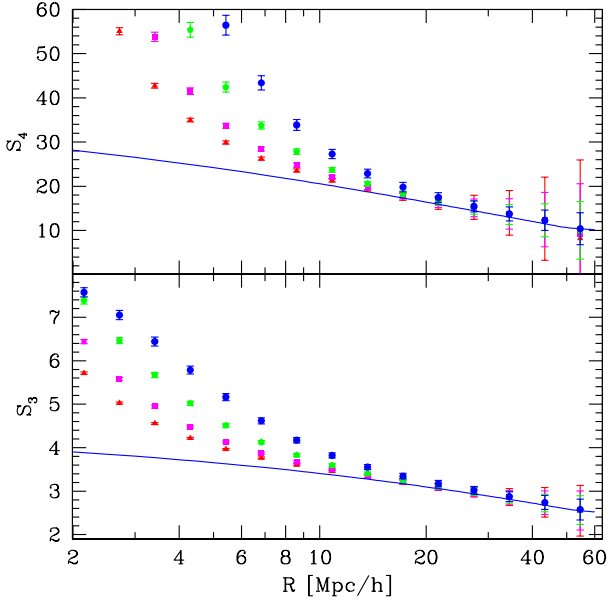}
\caption{The skewness $S_3$ (bottom) and kurtosis $S_4$ (top) for the reference runs (2LPT $z_i=49$) at redshifts $z=0,1,2,3$ (circles, pentagons, squares, and triangles, respectively, from top to bottom in each panel.}
\label{S3S4ref}
\end{figure}

Figure~\ref{S3S4ref} shows the measurement of $S_3$ and $S_4$ as a function of smoothing scale $R$ for the reference runs HR2LPT49, after averaging over 20 realizations. The agreement at large scales with the perturbative prediction (solid lines) is excellent, better than 1\% for $S_3$ for $R>40 \Mpc$. Note that all measurements at different redshifts $z=0,1,2,3$ lie on top of each other in the large-scale limit as predicted by tree-level PT \citep{JBC93,B94}, and depart from it at increasingly large scales as $z$ decreases as expected from one-loop PT \citep{SF96,FoGa98}. Previous measurements in large-volume simulations such as the Hubble volume \citep{CSJC00} could not achieve such a clean measure of higher-order cumulants independent of $z$ due to transients effects from $z_i=35$ ZA initial conditions, as will become clear shortly.

\begin{figure}
\includegraphics[width=84mm]{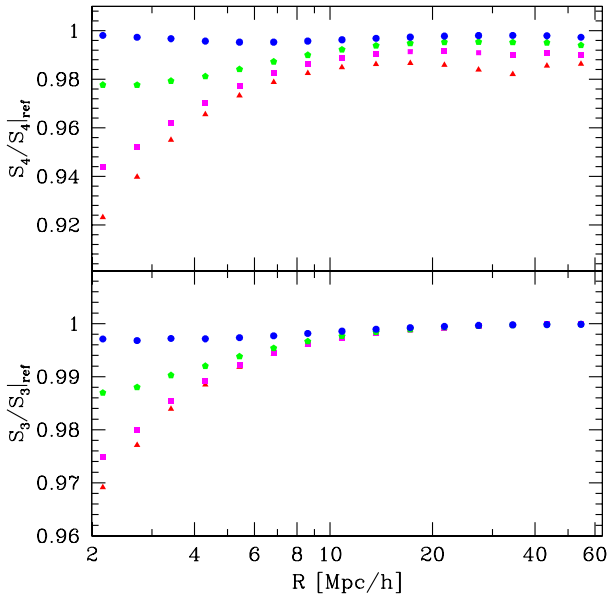}
\caption{The skewness $S_3$ (bottom) and kurtosis $S_4$ (top) for the 2LPT $z_i=11.5$ initial conditions compared to the reference runs ($z_i=49$) at redshifts $z=0,1,2,3$ (circles, pentagons, squares, and triangles, respectively, from top to bottom in each panel).}
\label{Sp11p5}
\end{figure}

\begin{figure}
\includegraphics[width=84mm]{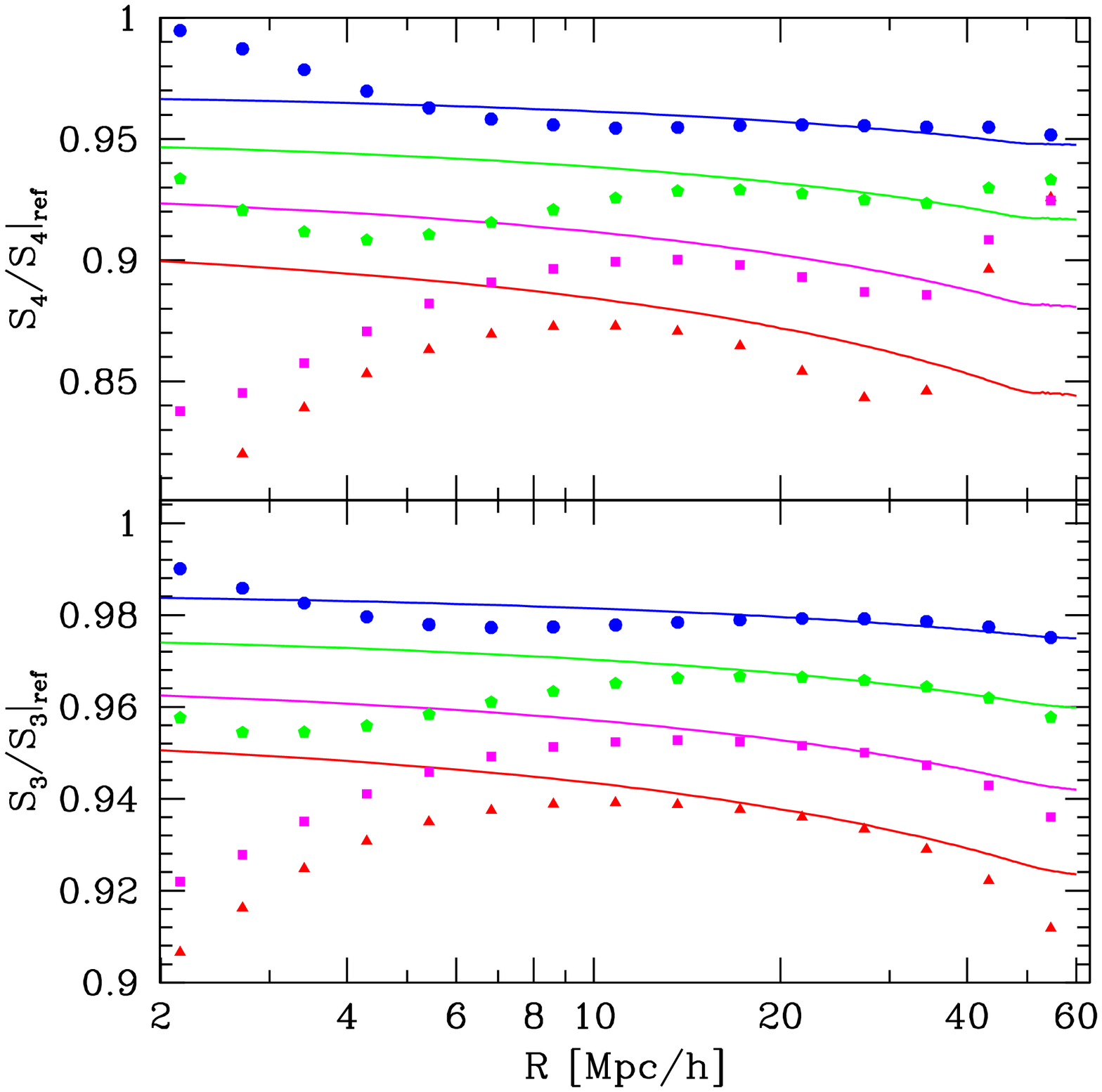}
\caption{Same as Fig.~\protect\ref{Sp11p5} but for ZA $z_i=24$ initial conditions. The solid lines show the predictions of the expected transient behavior in $S_3$, Eq.~(\protect\ref{S3trans}), and $S_4$, Eq.~(\protect\ref{S4trans}).}
\label{SpZA24}
\end{figure}

\begin{figure}
\includegraphics[width=84mm]{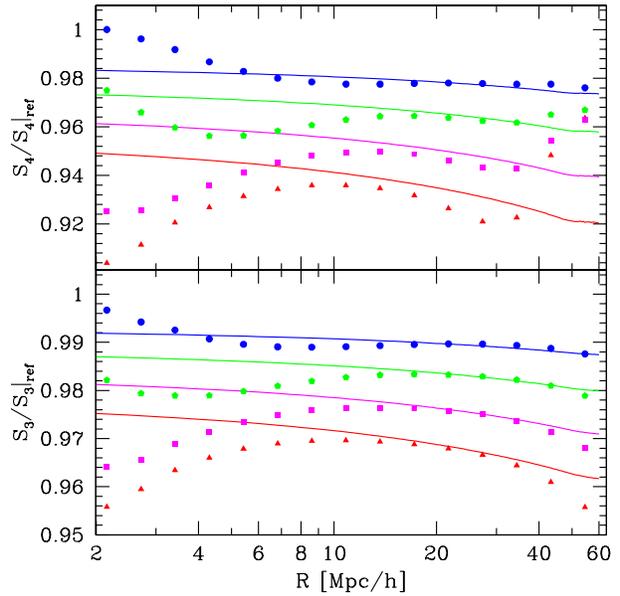}
\caption{Same as Fig.~\protect\ref{Sp11p5} but for ZA $z_i=49$ initial conditions. The solid lines show the predictions of the expected transient behavior in $S_3$, Eq.~(\protect\ref{S3trans}), and $S_4$, Eq.~(\protect\ref{S4trans}).}
\label{SpZA49}
\end{figure}

Figure~\ref{Sp11p5} shows what happens with 2LPT  initial conditions when the starting redshift is changed to a very late start, $z_i=11.5$. As expected theoretically, at large scales there should be no change whatsoever in $S_3$ since 2LPT reproduces exactly the second-order growing modes, and that's indeed what is seen in the bottom panel. At third-order, 2LPT does underestimate $S_4$ slightly, and that manifests in a small transient behavior at large scales (top panel). Note that since third- and higher-order couplings are slightly smaller in 2LPT, one expects the loop corrections to $S_3$ and $S_4$ to be smaller than in the reference runs, that's indeed what is seen in both panels in Fig.~\ref{Sp11p5}  at small scales: there is a slight underestimate of the small-scale cumulants, although by the time one reaches $z=0$, that is well under 1\%. Overall, the behavior of 2LPT initial conditions is remarkably stable given the very late start, $z_i=11.5$. In fact, we picked such a late start to make the effects visible at all at $z\la 3$.

Figures~\ref{SpZA24} and~\ref{SpZA49} show the corresponding situation with ZA initial conditions for $z_i=24,49$, respectively. It is obvious from these figures that the ZA initial conditions runs are not stable at large scales, in particular, $z_i=24$ and $z_i=49$ ZA runs do not give the same results, with significant deviations at the few percent level. These deviations are in excellent agreement with the PT predictions, Eq.~(\ref{S3trans}) for the skewness (bottom panels) and Eq.~(\ref{S4trans}) for the kurtosis (top panels). From this we conclude that the simulations started with ZA initial conditions have transients which are well understood at large scales from first principles. Figs.~\ref{SpZA24}-\ref{SpZA49} check the PT transients predictions more accurately than done before \citep{Sco98}.
We now discuss how these effects impact on clustering at small scales.

\subsection{Power Spectrum}
\label{secPower}

\begin{figure}
\includegraphics[width=84mm]{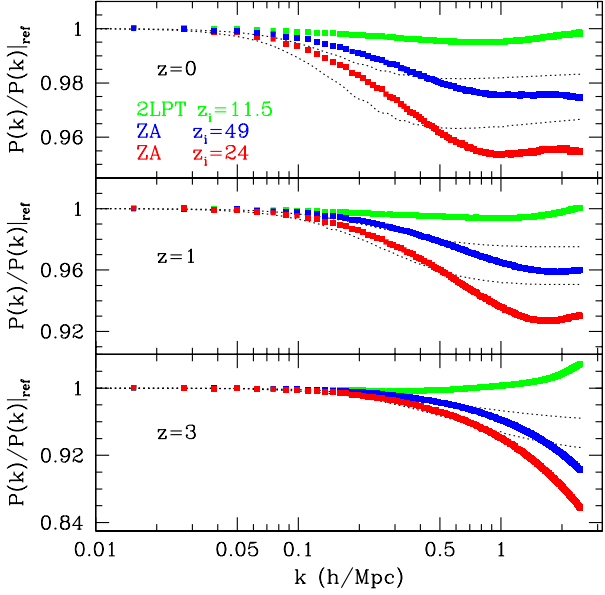}
\caption{Power spectrum for different initial conditions (2LPT $z_i=11.5$, ZA $z_i=49$ and ZA $z_i=24$, from top to bottom in each panel) compared to the reference runs at $z=0$ (top), $z=1$ (middle) and $z=3$ (bottom). The dotted lines show an estimate of the transients for ZA $z_i=49$ and ZA $z_i=24$ from one-loop PT.}
\label{PkReal}
\end{figure}

\begin{figure}
\includegraphics[width=84mm]{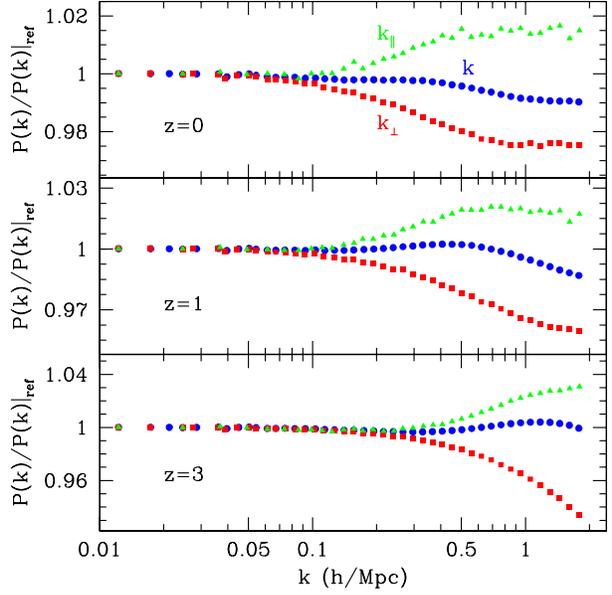}
\caption{Redshift-space power spectrum for ZA $z_i=49$ initial conditions compared to the reference runs at $z=0$ (top), $z=1$ (middle) and $z=3$ (bottom). In each panel we show modes parallel to the line of sight (triangles), monopole (circles) and modes perpendicular to the line of sight (squares).}
\label{PkRed}
\end{figure}

The power spectrum is the most widely used statistic. Nonlinear corrections to the power spectrum are controlled by  the same nonlinear couplings that determine higher-order statistics at large scales, and thus we expect to see transient behavior in the nonlinear power spectrum as well.

Figure~\ref{PkReal} shows that this is indeed the case. First, in each panel the top line denotes the 2LPT $z_i=11.5$ initial conditions runs, showing that in the 2LPT case there is almost no transient behavior, as expected from the results on the $S_p$ parameters at large scales. On the other hand, the ZA initial conditions runs show significant transient effects, up to $10\%$ at $z=3$, which can potentially lead to serious systematic errors for precision studies of the high-redshift universe. For example, a ZA start at $z_i=30$ was used for the LCDM simulation from which the latest fitting formula for the  nonlinear power spectrum \citep{Smith03} was derived. Our reference runs show a power spectrum for $k \la 2 \kMpc$ that is as much as $7\%$ higher  at $z=0$ and as much as $13\%$ higher at $z=3$ when compared to the \citet{Smith03} fitting formula. These deviations are in reasonable agreement with the expectations based on Fig.~\ref{PkReal}. A detailed comparison of the nonlinear power spectrum in our simulations to fitting formulae, one-loop PT and renormalized PT \citep{CS06} will be discussed elsewhere.

It is also important to note that a relatively high-redshift ZA start such as $z_i=149$, which we ran as a test,  leads only to mild improvements compared to ZA $z_i=49$ in Fig.~\ref{PkReal}, i.e. the suppression in the nonlinear regime is still of order $1\%$, consistent with the $a^{-1}$ slow scaling of transients in the ZA. We have checked that these results do not depend on the accuracy of the force, time integration, and softening. For example, using the HQ runs (see table~\ref{SimTable}) for $z_i=49$ we obtain very similar results to those in Fig.~\ref{PkReal}, specifically within about $0.04\%$ for $z=0,1$ and $0.02\%$ for $z=3$. Also, as shown in Fig.~\ref{PkReal}, the magnitude of the measured transients in the  ZA $z_i=49$ and ZA $z_i=24$ runs are in reasonable agreement with their estimation by one-loop PT using the kernels in Eq.~(\ref{recursionkernels}), particularly at high redshift where the agreement should be best. 

Figure~\ref{PkRed}  presents results in redshift space for the ZA $z_i=49$ initial conditions runs. The combined effect of density and velocity transients leads to nontrivial behavior. As shown by \citet{Sco98} transients in the velocity field are larger than in the density field. For modes parallel to the line of sight (triangles in each of the panels in Fig.~\ref{PkRed}), small-scale velocities suppress power due to velocity dispersion. Therefore, although transients in real space make the power smaller, when mapped into redshift-space with velocities that are even more suppressed by transients than the density, the overall effect is to {\em increase} the power spectrum along the line of sight compared to the reference runs that have no transients. Since there is cancellation between density and velocity transients, the overall effect is somewhat weaker than in the real space power. 

For modes perpendicular to the line of sight, on the other hand, the effect is opposite since these modes are unaffected by redshift-space distortions and show the same suppression in power seen in Fig.~\ref{PkReal}. This means that when 
the redshift space power spectrum is averaged over angles with respect to the line of sight to obtain the monopole, these opposite behaviors lead to cancellation that makes the  monopole less sensitive to transients (see circles in Fig.~\ref{PkRed}). However, the anisotropy of the redshift-space power spectrum is affected by transients much more than the monopole.

\subsection{Bispectrum}
\label{secBisp}

\begin{figure}
\includegraphics[width=84mm]{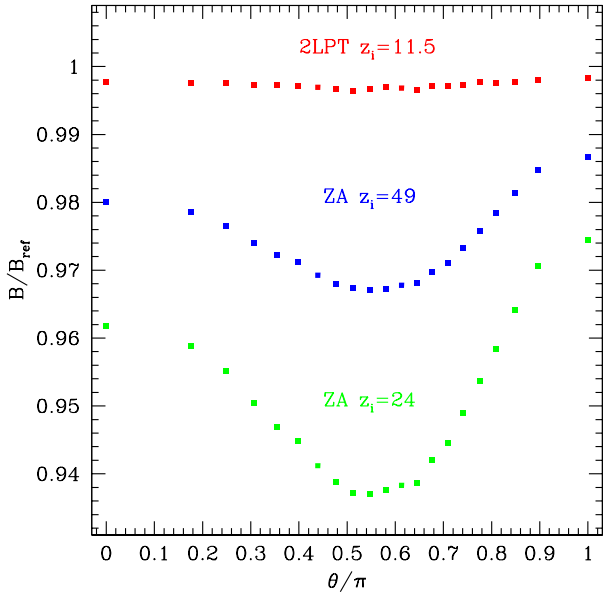}
\caption{Bispectrum for different initial conditions (2LPT $z_i=11.5$, ZA $z_i=49$ and ZA $z_i=24$, from top to bottom) compared to the reference runs at $z=0$. The triangles shown correspond to $k_1=0.12 \kMpc$, $k_2=2k_1$ with $\theta$ the angle between $\kv_1$ and $\kv_2$.}
\label{BispT}
\end{figure}

The bispectrum is also affected by transients, and from Eqs.~(\ref{F2trans}) and~(\ref{Bisp2E}) we see that at the largest scales where second-order Eulerian PT holds it should be affected most for triangles that are most different from elongated triangles (which correspond to $x=1$ and show no transient behavior). Figure~\ref{BispT} shows the bispectrum at $z=0$  for different initial conditions divided by that in the reference runs for triangles with $k_1=0.12 \kMpc$ and $k_2=2k_1$ as a function of  the angle $\theta$ between $\kv_1$ and $\kv_2$. From this we see again that the 2LPT $z_i=11.5$ initial conditions show very little ($\la 0.2 \%$) transient effects, whereas the ZA initial conditions runs show the expected transient signature, minimal at elongated triangles and maximal at isosceles triangles. These can lead to systematic errors in the determination of bias and cosmological parameters, given present expected observational errors \citep{SCPS06}.

\subsection{The PDF}
\label{secPDF}

\begin{figure}
\includegraphics[width=84mm]{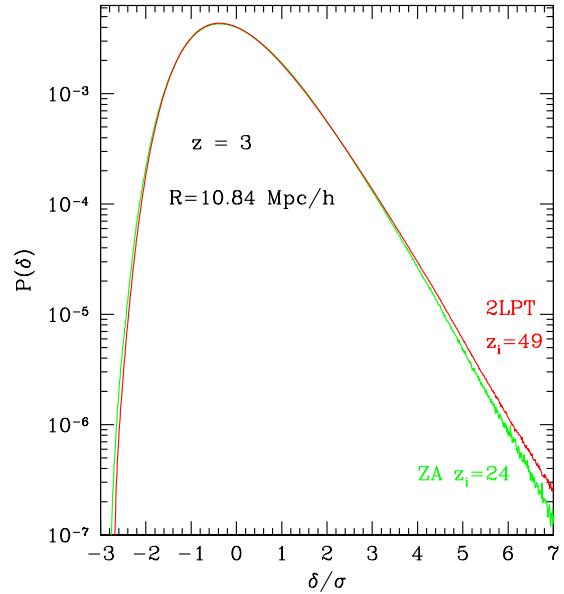}
\caption{The PDF of the density contrast $\delta$ in units of {\em rms} value, $\sigma \equiv \langle \delta^2 \rangle^{1/2} $, for our reference run (2LPT $z_i=49$) and ZA $z_i=24$ initial conditions.}
\label{PDFT}
\end{figure}

\begin{figure}
\includegraphics[width=84mm]{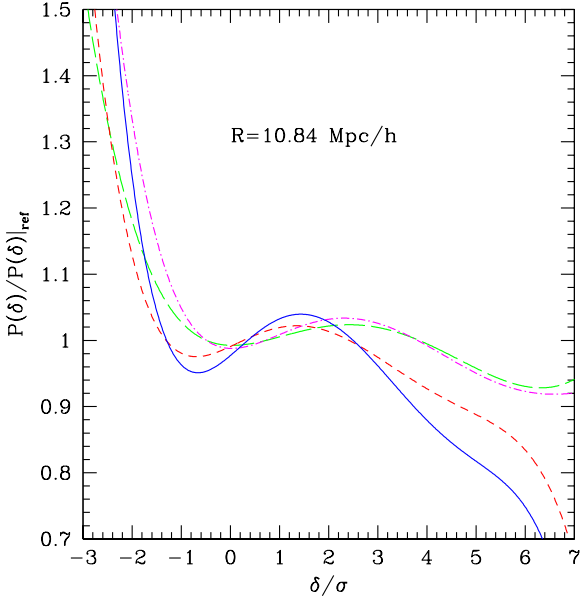}
\caption{Ratio of PDF's for different initial conditions to our reference run. Solid lines denote ZA $z_i=24$ at $z=3$, dashed lines ZA $z_i=49$ at $z=3$, dot-dashed ZA $z_i=24$ at $z=0$, long dashed lines ZA $z_i=49$ at $z=0$. }
\label{PDFTratio}
\end{figure}

The suppression of higher-order moments by transients means that the probability distribution function (PDF) of density fluctuations is also affected. Figure~\ref{PDFT} shows a comparison of the PDF of our reference run (2LPT $z_i=49$) and ZA $z_i=24$ initial conditions at redshift $z=3$ for a smoothing scale $R=10.84 \Mpc$. Note how the non-Gaussian features of the PDF are changed by transients: the right tail is suppressed, whereas the left cutoff is enhanced, leading to smaller skewness. Figure~\ref{PDFTratio} shows the ratio of  the PDF for different initial conditions to that of the reference run, to better appreciate the differences. Although the differences at the left tail are more significant, the PDF is falling very steeply, thus in practice the differences seen at the right tail are more important. These correspond to high-density regions, and thus we expect these differences to impact the high-mass tail of the mass function of dark matter halos.

\subsection{The Mass Function}
\label{secMF}

\begin{figure}
\includegraphics[width=84mm]{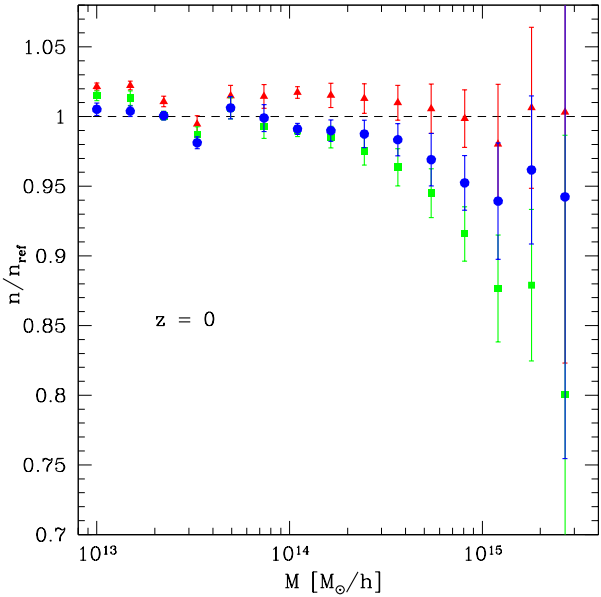}
\caption{Mass function for different initial conditions (2LPT $z_i=11.5$, ZA $z_i=49$ and ZA $z_i=24$, from top to bottom) compared to the reference runs at $z=0$.}
\label{massFz0}
\end{figure}

\begin{figure}
\includegraphics[width=84mm]{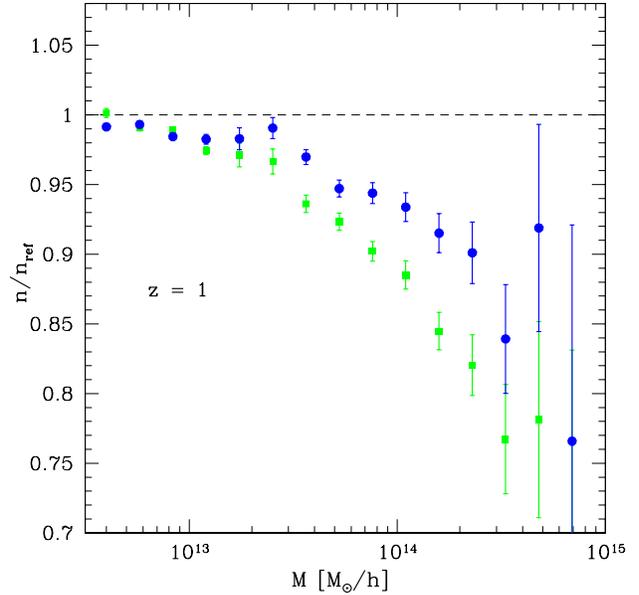}
\caption{Mass function for ZA initial conditions with $z_i=49$ (circles) and $z_i=24$ (squares) compared to the reference runs at $z=1$.}
\label{massFz1}
\end{figure}

Figures~\ref{massFz0} and~\ref{massFz1} show the ratio of mass functions for different initial conditions to the reference runs, where  the LR simulations have been used at the high-mass end to probe halos with masses larger than $m\simeq 10^{14}\, M_\odot/h$. All masses have been corrected by the procedure described in \citet{WAHT05} to take into account the finite number of particles making up halos. We use the friends-of-friends algorithm with linking length parameter equal to $0.2$ to identify halos.

The effect in the suppression of the high-mass tail is  clearly seen in the simulations that were started with ZA initial conditions. 
The mass function in our reference runs is 15\% larger than the \citet{WAHT05} fitting formula for $m=10^{15}h^{-1}M_\odot$ at $z=0$ (and thus about $30\%$ larger than the \citet{ST99} mass function), whereas for $m<10^{14}h^{-1}M_\odot$ is consistent with the \citet{WAHT05} fit. The \citet{Jetal01} fit is very similar to the  \citet{WAHT05} fit in the mass range covered by our simulations. These differences, in good agreement with the deviations seen in Fig.~\ref{massFz0}, are thus most likely due to transients in the simulations used to derived these fitting formulae at the high-mass end, which were started from $z_i=24,35$ ZA initial conditions.

Such a suppression of the mass function is clearly important for current and future observations that use the abundance of massive clusters to probe cosmology. It is also worth noting that the suppression of the mass function high-mass tail due to transients can be made worse by requiring that simulations start at a redshift $z_i$ such that density fluctuations at the interparticle distance be a fixed number (often chosen to be about $0.2$). This means that the large volume simulations that are required to probe rare events will be started later and thus will be subject to more suppression due to transients.

Using the LRs simulations with smaller softening length  (see table~\ref{SimTable}) for $z_i=49$ we obtain very similar results to those in Fig.~\ref{massFz0} for $z=0$, specifically the average deviation in the eight bins with $m>10^{14} M_\odot/h$ (excluding the last one that has substantial errors) is within $0.7\%$ of that in Fig.~\ref{massFz0}. Therefore the suppression seen in the mass function is robust to a rather large change in softening length.

\citet{Retal03} studied the effects of the starting redshift $z_i$ on the mass function measured at high redshifts $z\simeq7-15$ from numerical simulations initialized with the ZA. Their statistical errors are rather large, but they do seem to see the effects we report here when comparing simulations with $z_i=69,139,279$. They conclude $z_i=139$ is safe for measurements at $z\simeq7-15$ and find a suppression of the high-mass tail from the  \citet{ST99} mass function which becomes stronger with $z$. Extrapolating our results to their range, we would expect an important contribution from transients for those choices of  $\{z,z_i\}$ going exactly in the same direction with increasing $z$. Therefore we suggest that high-redshift studies of the high-mass tail of the mass function should be carefully checked for transients.

\subsection{Halo Bias}
\label{secHB}

\begin{figure}
\includegraphics[width=84mm]{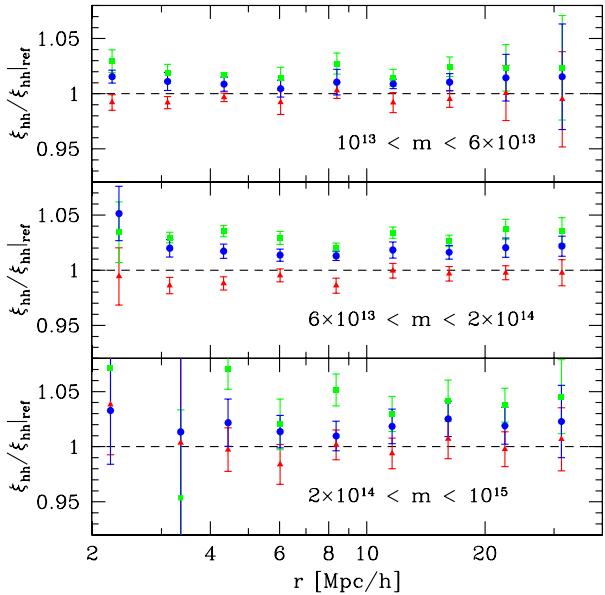}
\caption{Ratio of halo-halo correlation functions $\xi_{\rm hh}$ at $z=0$ for different initial conditions, 2LPT $z_i=11.5$ (triangles), ZA $z_i=49$ (circles) and ZA $z_i=24$ (squares), compared to the reference runs at $z=0$. The different panels correspond to three bins in halo mass as indicated (masses are in units of $h^{-1}M_\odot$).}
\label{HBias0}
\end{figure}

Finally, Fig.~\ref{HBias0} shows how the clustering of dark matter halos is affected by transients at $z=0$. The panels show the halo-halo correlation function $\xi_{\rm hh}$ for different initial conditions compared to those in the reference runs, for three different mass bins. Here we have used HR simulations for the first bin in mass, and the LR simulations for the two largest mass bins. Note that as halo mass becomes larger, corresponding to rarer events, the halo bias in the simulations with more transients is {\em larger}. This is consistent with the fact that transients suppress the abundance of rare halos making them more rare and thus more biased, as expected from the peak-background split.

\section{Summary and Conclusions}
\label{Conc}

In this paper we considered the use of different approximations to set up initial conditions for cosmological simulations, concentrating on typical initial redshifts $z_i \simeq 50$, and studied their influence on low-redshift ($z \la 3$) statistics. The standard procedure in the literature is to use the \citet{ZA70} approximation (ZA), a linear solution in Lagrangian space, or even linear perturbation theory in Eulerian space (typically when simulating baryons using methods which are not based on particles). 

At high redshifts ($z \sim 50$) the naive expectation is that linear perturbation theories should be accurate enough, but they are not. The reason for this is that at high redshifts nonlinear corrections are relatively much more important than at low redshift  because the relevant scales involved have a very steep spectrum (with the effective spectral index very close to $n_{\rm eff}=-3$). An example of this is shown in Fig.~\ref{PDFT}, where the density PDF smoothed on $10 \Mpc$ at $z=3$ shows strong non-Gaussian features, even though naively these scales are in the linear regime (since the linear {\em rms} density fluctuation is only $\sigma =0.25$). Modeling these nonlinear corrections using the ZA leads to a significant underestimate of nonlinear couplings and thus non-Gaussianity. This leads to a suppression of the density PDF tail for large densities that translates into a suppression of the high-mass tail of the mass function. The weaker nonlinear couplings lead to a smaller power spectrum in the nonlinear regime.

The mismatch between the approximation used to set initial conditions and the correct dynamics leads to {\em transients}, excitation of decaying modes that are long-lived and survive until low redshift, systematically biasing low the nonlinear couplings even after evolving with the correct dynamics of an N-body code for as much as 50-100 scale factors. Transients cannot be avoided, but they can be significantly reduced by imposing initial conditions with better approximations to the nonlinear equations of motion.

More specifically, in this paper we have shown that,

\begin{enumerate}

\item The measurements of statistics at low redshift for simulations started with ZA initial conditions depend significantly on the starting redshift $z_i$, signaling the presence of transients.

\item This can be avoided to a large extent by using  2LPT initial conditions, which as we have demonstrated show very little dependence on $z_i$ even for extremely late starts such as $z_i=11.5$. Given that such runs show at most 1\% transient effects and that transients in 2LPT decay as inverse square of the scale factor, we can estimate that our reference runs with $z_i=49$ are accurate (in terms of transients) to better than a tenth of a percent for $z\la 3$.

\item The simplest way to quantify transients is to follow how non-Gaussianity develops at large scales as nonlinear couplings reach their full strength during the transient evolution. Our results for higher-order cumulants at large scales confirm previous PT results \citep{Sco98,Val02} with much better accuracy than done before.

\item The  small scale behavior of transients is consistent with the reduction of nonlinear couplings observed at large scales, and we have checked it is not affected by the choice of softening length, force and time integration accuracy.

\item The use of the ZA at $z_i=49$ leads to suppressions in the power spectrum of order a few ($z=0$) to almost ten percent ($z=3$) in the nonlinear regime. These results are in reasonable agreement with an estimation of transients in the power spectrum based on one-loop PT.

\item  ZA initial conditions also lead  to a suppression in the high-mass tail of the mass function of dark matter halos. For $z_i=24$ this suppression is about 10\% for $m=10^{15}h^{-1}M_\odot$ at $z=0$ and $m=10^{14}h^{-1}M_\odot$ at $z=1$. The clustering of halos is also affected at the few-percent level. 

\end{enumerate}

An obvious alternative to using 2LPT initial conditions is to start with the ZA much earlier than $z_i\simeq 50$. However, in order to compete with the improvement brought by using 2LPT very high redshifts are required. For example, Fig.~\ref{PkReal} shows that a $z_i=11.5$ 2LPT start has approximately a 0.5\% transient effect on the nonlinear power spectrum at $z=0$, whereas a $z_i=49$ ZA start has a 2.5\% effect. In order for the ZA to reduce the transients to the level of the 2LPT $z_i=11.5$ start, a $z_i=249$ ZA start would be required. Reaching the level of a 2LPT $z_i=49$ start, corresponding to our reference runs, will require about another factor of 16 in redshift (given that transients scale as $a^{-2}$ in 2LPT compared to $a^{-1}$ in the ZA). Other issues make starting simulations at very high redshifts a bit more challenging, e.g.  depending on the N-body code algorithm, numerical errors may be more difficult to avoid at very high redshifts where perturbations are so small. In addition, initial conditions for baryons at such high redshifts must be carefully included as they are still falling into dark matter potential wells. 

Although we have not studied this in any detail, simulations started with linear theory instead of the ZA are expected to be much more significantly affected by transients, e.g. a simulation started with linear theory at $z_i=49$ should be roughly equivalent to one started with the ZA at $z=11.5$ (see Fig.~\ref{S3ZALT}). Note that linear theory initial conditions are often used in codes that rely on grid methods with gas dynamics solvers. We expect such methods to show differences with e.g. SPH methods that use ZA initial conditons. When the dynamical variables are densities and velocities in a mesh (instead of information carried by particles), a way to reduce transients would be to use higher-order Eulerian perturbation theory.

While in this paper we have concentrated on statistics of the dark matter density perturbations, it would be interesting to know how much do transients affect other high-redshift probes, where transients are most likely to play an important role. In this regard, the most pressing aspect  would be to check the impact of transients in the Lyman-$\alpha$ forest power spectrum, because of the high accuracy demanded from simulations to obtain cosmological constraints from current data sets \citep{VHL06,SSM06}. 

The use of better initial conditions should be a useful tool in reaching  future goals of simulating the nonlinear power spectrum 
to an accuracy of better than 1\% percent  needed for the next-generation of weak gravitational lensing surveys \citep{HuTa05}. 

The code we used to generate ZA \& 2LPT initial conditions in this paper is publically available at {\tt http://cosmo.nyu.edu/roman/2LPT}.

\section*{Acknowledgments}

We thank S.~Colombi, H.~Couchman,  S.~Habib, K.~Heitmann, L.~Hui, A.~Jenkins, A.~Klypin, A.~Kravtsov, U.~Seljak, R.~Smith, V.~Springel, R.~Thacker, M.~Warren, D.~Weinberg, for useful discussions and E.~Sefusatti, M.~Masjedi, and R.~Smith for comments on the manuscript. We thank NYU Information Technology Services for making  its High Performance Computation Cluster available to us, and thank Joseph Hargitai for his technical assistance. We acknowledge use of the University of Washington N-body shop friends-of-friends code to find dark matter halos.

\appendix

\section{Second-Order Lagrangian Perturbation Theory (2LPT)}
\label{sec2LPT}

Here we present the basics of 2LPT, we refer the reader to  the literature \citep{Buchert94,BMW94,BCHJ95,Catelan95, Sco00b} for more information. A detailed step by step implementation of 2LPT initial conditions for numerical simulations is presented in Appendix~D2 of \citet{Sco98}. Here we assume $\Omega_m=1$ and $\Omega_\Lambda=0$. For precision work one must take into account the cosmological dependence, including radiation as well. 

In Lagrangian dynamics, particle positions are described by a displacement field $\Psi$ so that,
\beq
\xv = \qv + \Psi,
\label{LagToEu}
\eeq
where particle trajectories obey the equation of motion,
\beq
\frac{d^2\xv}{d \tau^2}+{\cal H}(\tau)\, \frac{d\xv}{d \tau} = -\nabla \Phi,
\eeq
where $\Phi$ is the gravitational potential. This leads to the following equation of motion for the displacement field,
\beq
J(\qv,\tau)\, \nabla \cdot \Big[ \frac{d^2\xv}{d \tau^2}+{\cal H}(\tau)\, \frac{d\xv}{d \tau} \Big]= \frac{3}{2} {\cal H}^2 (J-1),
\label{EoMPsi}
\eeq
where we have used the Poisson equation and $1+\delta(\xv) =1/J$ where $J$ is the Jacobian of the mapping, 

\beq
J(\qv,\tau)={\rm Det}(\delta_{ij}+\Psi_{i,j}),
\label{Jac}
\eeq
and $\Psi_{i,j} \equiv \partial \Psi_i/\partial q_j$. 

Solving Eq.~(\ref{EoMPsi}) perturbatively leads to positions and velocities,
\beqa
\label{pos}
\xv &=& \qv - a\, \nabla_{\qv}\phi^{(1)} -\frac{3}{7}\, a^2\,  \nabla_{\qv}\phi^{(2)} \\
\label{vel}
\vv &=&  - a\, H\, \nabla_{\qv}\phi^{(1)} -\frac{6}{7}\, a^2\,  H\, \nabla_{\qv}\phi^{(2)},
\eeqa
where the potentials satisfy the Poisson equations  \citep{BMW94},
\beqa
\nabla_{\qv}^2\phi^{(1)}(\qv) &=& \delta(\qv) \\
\nabla_{\qv}^2\phi^{(2)}(\qv) &=& \sum_{i>j} \Big\{ \phi^{(1)}_{ii}(\qv)  \phi^{(1)}_{jj}(\qv) - [ \phi^{(1)}_{ij}(\qv)]^2 \Big\}.
\eeqa

Setting $\phi^{(2)}=0$ in Equations~(\ref{pos}-\ref{vel}) leads to the \citet{ZA70} approximation.

Going beyond 2LPT to third-order (3LPT) becomes more costly due to the need to solve three additional Poisson equations  \citep{BMW94,Catelan95}, and the improvement in higher-order statistics is modest \citep{Sco00b}, leading to an improvement by only a factor of two in the scale factor needed to suppress transients \citep{Sco98}, whereas going from ZA to 2LPT the  improvement is more than one order of magnitude. However, 3LPT does provide a better behavior in underdense regions \citep{BCHJ95}, and may be worth considering for studies sensitive to the statistical properties of voids. 

\section{On Glass initial conditions}
\label{secGlass}

\begin{figure}
%\vspace{174pt}
\includegraphics[width=84mm]{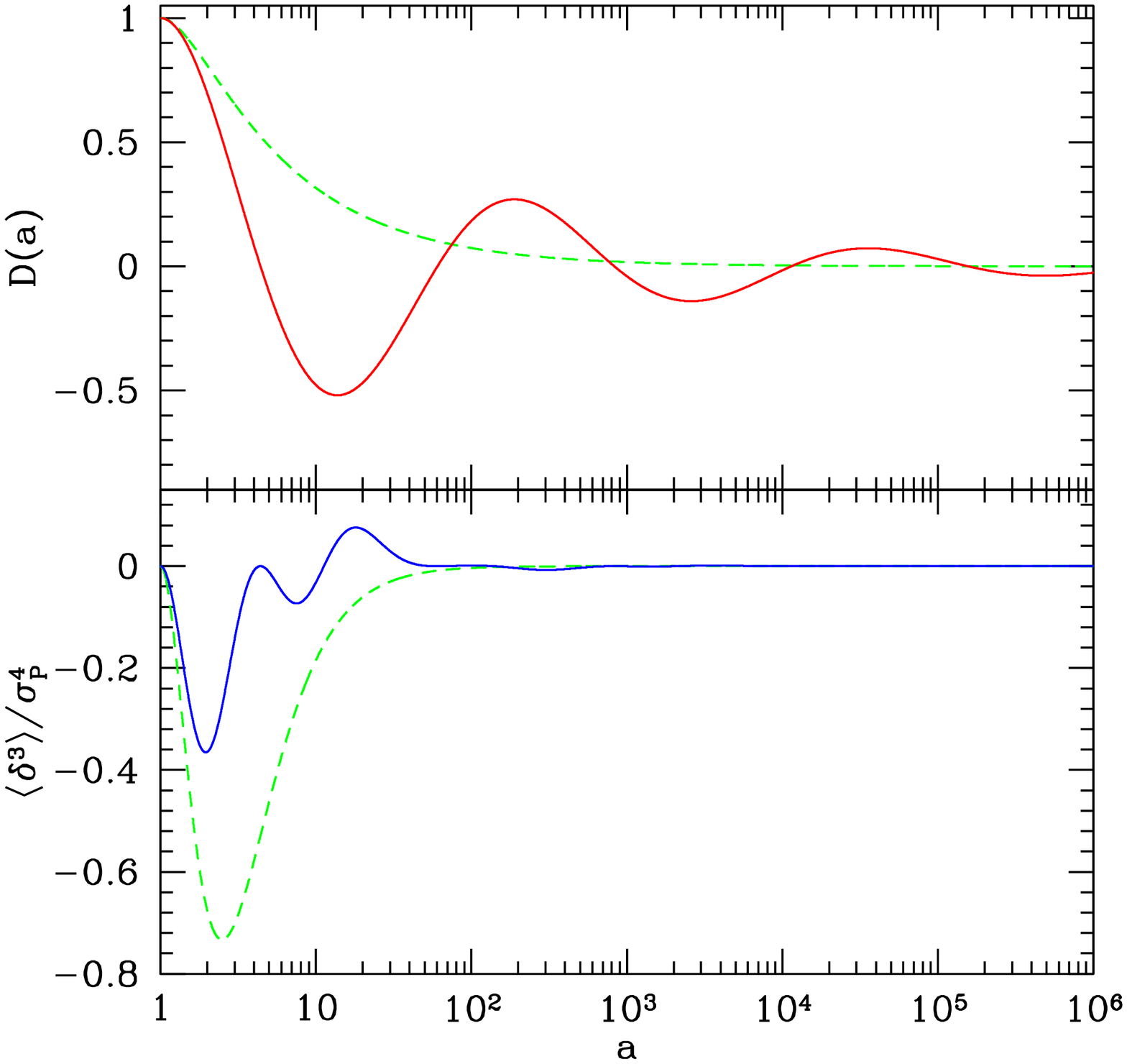}
\caption{Evolution of perturbations towards a glass. The top panel shows the growth factor $D$ as a function of scale factor $a$ for standard procedure (solid) and extra-damped procedure with $\nu=3$ (dashed). The bottom panel shows the corresponding behavior of the third-moment in terms of its Poisson value.}
\label{FigGlass}
\end{figure}

In this paper, our numerical simulations were started from a grid. An alternative to a grid start is to displace particles from a configuration known in the literature as a glass, which is obtained starting from a Poisson distribution and running an N-body code with the sign of the acceleration reversed \citep{CaCo89,BGE95,CTP95,White96,Smith03}. 

In this appendix we derive analytically how perturbations evolve towards a glass, and what steps must be taken to avoid exacerbating the problem of transients. Reversing accelerations means that the equations of motion are still given by Eq.~(\ref{eom}) but with 
\beq
\label{OmabG}
\Omega_{ab} \equiv \Bigg[ 
\begin{array}{cc}
0 & -1 \\ 3/2 & 1/2 
\end{array}        \Bigg],
\eeq
instead of Eq.~(\ref{Omab}). As a result the linear propagator, Eq.~(\ref{propdef}), reads 
\beqa
g_{ab}(\eta) & = & {\rm e}^{-\eta/4}\ \Bigg\{\cos\Big(\frac{\sqrt{23}}{4}\eta\Big) \, 
\Big[ \begin{array}{rr} 1 & 0 \\ 0 & 1 \end{array} \Big] \nonumber \\ & & +
\frac{1}{\sqrt{23}} \sin\Big(\frac{\sqrt{23}}{4}\eta\Big) \, 
\Big[ \begin{array}{rr} 1 & 4 \\ -6 & -1 \end{array} \Big] \Bigg\},
\label{propG}
\eeqa 
instead of Eq.~(\ref{prop}). The ``growth" factor $D$ for density perturbations can be obtained by specifying how initial conditions are set before evolving towards a glass. Assuming the common choice of zero initial velocities, we have $D= g_{1b}\cdot (1,0)$ and then

\beq
D = \frac{1}{a^{1/4}}\ \Big[ \cos\Big(\frac{\sqrt{23}}{4}\ln a\Big) +
\frac{1}{\sqrt{23}} \sin\Big(\frac{\sqrt{23}}{4}\ln a\Big) \Big],
\label{growthG}
\eeq
where we have switched to the scale factor as a time variable (recall $\eta = \ln a$). Therefore, both eigenmodes contribute an overall decaying factor of $a^{-1/4}$ with oscillations of opposite phase. The top panel in Fig.~\ref{FigGlass} shows Eq.~(\ref{growthG}) as a function of $a$ (solid), showing that the decay of perturbations due to repulsive gravity is very slow. These damped oscillations explain the behavior seen in Fig.~A2 in \citet{BGE95}. 

Using Eq.~(\ref{recursionkernels}) with the propagator in Eq.~(\ref{propG}) and Gaussian Poisson initial conditions we can derive the behavior of the third moment as evolution proceeds towards a glass,

\beqa
\langle \delta^3 \rangle_{\rm glass} &=& -\frac{3D^2}{299\sqrt{a}} \Big[
52+ (23 a^{1/4}-75)\,  \cos\Big(\frac{\sqrt{23}}{4}\ln a\Big) \nonumber \\ & & 
+ \sqrt{23} \, (17 a^{1/4}-9)\, \sin\Big(\frac{\sqrt{23}}{4}\ln a\Big) 
\Big] \ \sigma_{\rm P}^4,
\label{cum3G}
\eeqa
where $\sigma_{\rm P}^2 $ denotes the initial Poisson variance. This result is shown as a function of $a$ in the bottom panel in Fig.~\ref{FigGlass}  (solid). Note that (mostly negative) skewness is developed as evolution to the glass takes place, therefore although the amplitude of perturbations mostly decreases as things evolve to a glass, {\em skewness is generated}. Unless one evolves by many ($\ga 10^2$) scale factors this can potentially enhance the problem of transients. The reason is that if the glass itself has negative skewness, this adds to the incorrect skewness that initial conditions may have, as we now discuss. This issue does not arise for grid initial conditions.

Particles are displaced from their glass positions $\qv$ by the displacement vector $\Psi$ so that their positions are given by Eq.~(\ref{LagToEu}), and the induced density perturbations then obey
\beq
(1+\delta(\qv))\, d^3 q = (1+\delta(\xv))\, d^3 x,
\eeq
where $\delta(\qv)$ are the density perturbations due to the distribution of starting positions. These vanish for a grid below the Nyquist frequency, but are non-zero for a glass. Then, linearizing, the total perturbation imposed on a set of particles is 
\beq
\delta(\xv) = \Big(\frac{1}{J}-1\Big) + \delta(\qv) \equiv \delta_{\rm ic}(\xv) + \delta_{\rm glass}(\xv),
\label{delG}
\eeq
where the first term is the standard initial condition perturbation generated by the displacement field, with $J$ given by Eq.~(\ref{Jac}), and the second term is the glass perturbation, the outcome of the process described by Eqs.~(\ref{propG}-\ref{cum3G}). Since the two perturbations are independent, requiring that the glass skewness is negligible (and thus has no impact on transients) means to require it to be small compared to the Poisson noise of the particles  (which typically is smaller than the physical perturbations $\delta_{\rm ic}$ imposed by the initial condtions). Then one must simply require

\beq
\langle \delta^3 \rangle_{\rm glass} \ll \sigma_{\rm P}^4,
\eeq
which according to Fig.~\ref{FigGlass} requires  $a \ga 10^2$ in the evolution towards a glass. 

In the desire for faster convergence towards a glass, the Hubble drag term in the equations of motion is sometimes increased to avoid oscillations \citep{CTP95}. Making such a modification in Eq.~(\ref{OmabG}) to allow for a generic drag of magnitude $\nu$ gives, 
\beq
\label{OmabG2}
\Omega_{ab} \equiv \Bigg[ 
\begin{array}{cc}
0 & -1 \\ 3/2 & \nu 
\end{array}        \Bigg],
\eeq
which leads to propagator eigenmodes time dependence as ${\rm e}^{w_\pm \eta}$ with $2\, w_\pm =-\nu \pm \sqrt{\nu^2-6}$, and thus the requirement of no oscillations in approaching a glass gives $\nu^2 \geq 6$. Taking $\nu=3$ as an example, we obtain the growth factor and third moment as before, shown in Fig.~\ref{FigGlass} as dashed lines. We see from this that although the convergence to a glass is improved in the growth factor sense, the generation of skewness is more significant, and therefore one must wait roughly the same as with $\nu=1/2$ to have negligible impact. 

Summarizing, generating a glass with an overall expansion factor by $a\ga 10^2$ should be a safe configuration of points for starting cosmological simulations.

\bsp

\label{lastpage}

\end{document}